\documentclass[twocolumn]{article}
\input{epsf.sty}

 \epsfverbosetrue

\title{MAGNETIC FIELDS OF AGNS AND STANDARD ACCRETION DISK MODEL:
 TESTING BY OPTICAL POLARIMETRY\\ {\small (Accepted in Astronomy \& Astrophysics)}}

\author{N.A. Silant'ev$^{1,2}$\thanks{E-mail: silant@inaoep.mx}\,\,,
M.Yu. Piotrovich$^2$, Yu.N. Gnedin$^2$, T.M. Natsvlishvili$^2$\\
\medskip\\ (1) Instituto Nacional de Astrof\'isica, \'Optica y
Electr\'onica, Luis Enrique Erro 1,\\ Apartado Postal 51 y 216,
72840, Tonantzintla, Puebla, M\'exico\\ (2) Central Astronomical
Observatory at Pulkovo of Russian Academy of Sciences,\\ 196140,
Saint-Petersburg, Pulkovskoe shosse 65, Russia}

\begin{document}

\maketitle

\begin{abstract}
We have developed the method that allows us to estimate the
magnetic field strength at the horizon of a supermassive black
hole (SMBH) through the observed polarization of optical emission
of the accreting disk surrounding SMBH. The known asymptotic
formulae for the Stokes parameters of outgoing radiation are
azimuthal averaged, which corresponds to an observation of the
disk as a whole. We consider two models of the embedding
3D-magnetic field, the regular field, and the regular field with
an additional chaotic (turbulent) component. It is shown that the
second model is preferable for estimating the magnetic field in
NGC 4258. For estimations we used the standard accretion disk
model assuming that the same power-law dependence of the magnetic
field follows from the range of the optical emission down to the
horizon. The observed optical polarization from NGC 4258 allowed
us to find the values $10^3 - 10^4$ Gauss at the horizon,
depending on the particular choice of the model parameters. We
also discuss the wavelength dependencies of the light
polarization, and possibly applying them for a more realistic
choice of accretion disk parameters.

\end{abstract}

\section{Introduction}

 It is now commonly accepted that active galactic nuclei (AGNs)
and quasars (QSOs) frequently possess the magnetized accretion disks
(see, for example, the reviews of Blaes 2003; Moran 2008 on NGC 4258).
There are many models of
the accretion disk structures (see Pariew et al. 2003, and references therein).
The best known and most frequently used is
the standard model of Shakura \& Sunyaev (1973). The polarimetric
observations frequently demonstrate that AGNs and QSOs have polarized
emission in different wavelength ranges, from  ultraviolet to radio waves,
in continuum and in the line emission (see Martin et al. 1983;
Webb et al. 1993; Impey et al. 1995; Wilkes et al. 1995;
Barth et al. 1999; Smith et al. 2002; Modjaz et al. 2005). These papers
discuss the different mechanisms for the origin of the observed polarization:
 the light scattering in accretion disks, which happens on both free and bound
electrons, synchrotron radiation of charged particles. These  mechanisms
can work in different structures such as the plane and warped accretion disks and
toroidal clumpy rings, surrounding the accretion disks and jets. Frequently different models
 are proposed to explain the same source. There are a lot of
papers devoted to different aspects of the structure and emission of AGNs and
QSOs. Many theoretical papers propose the possible behavior of a magnetic field
in these objects.

In this paper we develop the technique of estimating the magnetic fields in
different parts of plasma accretion disks. Especially interesting is the estimation
of magnetic field in the horizon of the supermassive black holes in AGNs.
The  main idea is to use the observed
integral polarization from magnetized plasma accretion disks.
We use the known fact that the Faraday rotation of polarization plane changes both
the values of integral polarization degree $p$ and position angle $\chi $.
The  observed spectra $p(\lambda)$ and $\chi(\lambda)$ acquire very specific forms
due to Faraday rotation.
The detailed discussion and calculations of these effects are presented
in Silant'ev (1994), Dolginov et al. (1995), Gnedin \& Silant'ev (1997),
Agol and Blaes (1996), etc. The observed polarization possesses the information about
the magnetic field in magnetized electron atmospheres and can serve for estimating the field.
In Gnedin et al. (2006) the method was considered for pure vertical
magnetic field ${\bf B}_{\|}$ and without the correct azimuthal averaging of
the asymptotic formulae (4).

For estimating  the magnetic field we use the simple approximate formulae (Silant'ev 2002)
that represent solutions to a number of ``standard" problems of the radiative transfer theory
in magnetized electron atmospheres, namely, the Milne problem and the cases when the sources
of thermal radiation are distributed homogeneously, linearly, and exponentially in an
optically thick atmosphere. These ``standard" solutions allow us to approximate
the solution of problem with a more complex distribution of thermal sources inside
the atmosphere, because  the latter can be presented as a superposition of ``standard" sources.

\begin{figure}
\centering\leavevmode \epsfxsize=0.45\textwidth
\epsfbox{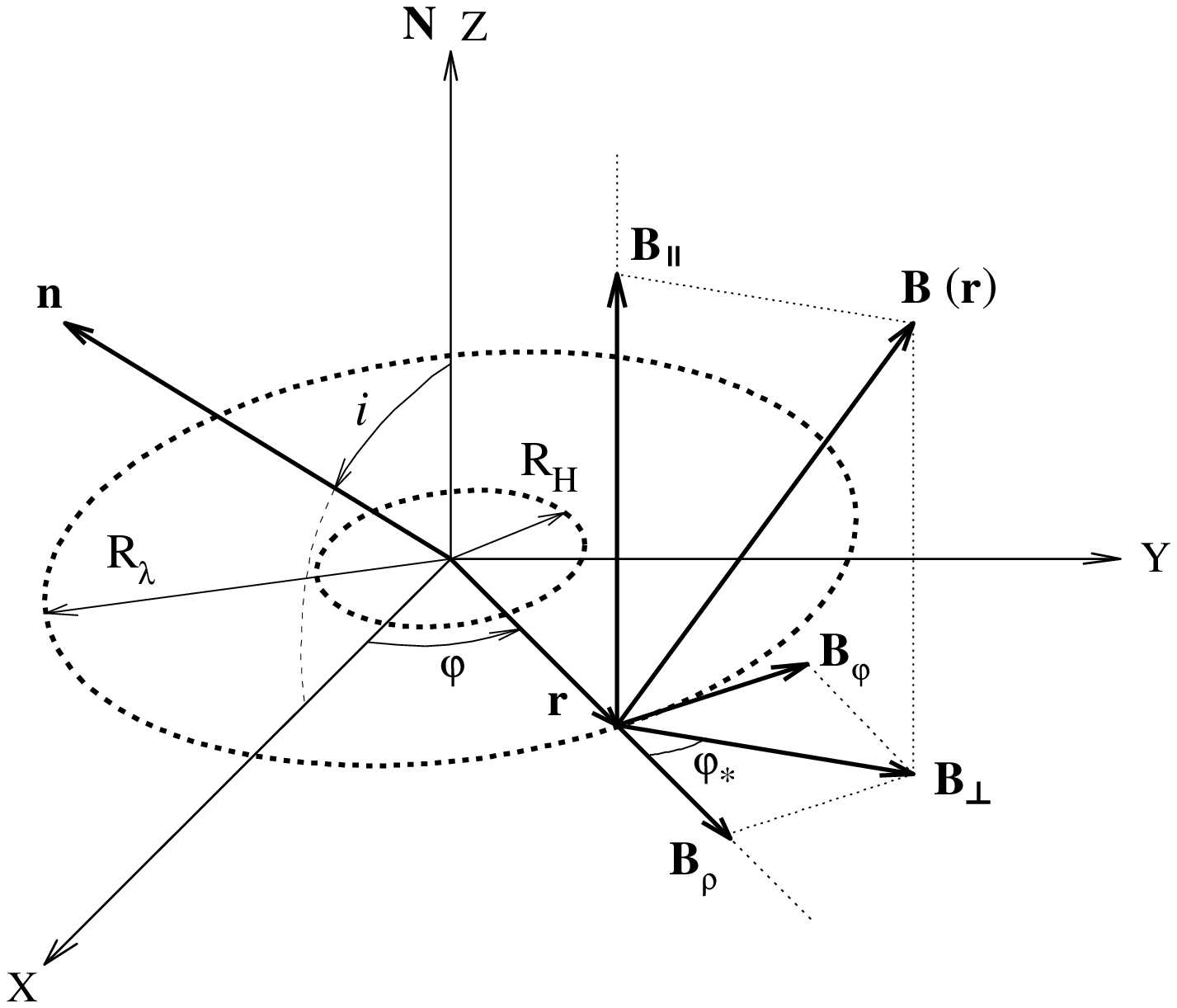} \caption{\small Main notions of accretion
disk surface, geometry of magnetic field ${\bf B}$, and line of
sight ${\bf n}$ (see text).} \label{fig1}
\end{figure}

For the optically thick accretion disks, the solution of the Milne
problem is used, i.e. the case  where the sources of thermal radiation are located far from
the surface.
The polarization and angular intensity distribution of outgoing radiation for non-magnetized
electron atmosphere is presented in the known Chandrasekhar's book (see Chandrasekhar 1950).
The numerical solution to Milne's problem for a magnetized electron atmosphere with
the  magnetic field ${\bf B}_{\|}$ parallel to the normal ${\bf N}$ to an atmosphere
is presented in Agol \& Blaes (1996), and Shternin et al. (2003).
The numerical solution
to this problem for the turbulent magnetized atmosphere is given in Silant'ev (2007).
In this paper the approximate formulae for Milne's problem are also generalized for
the case of a turbulent atmosphere. It is very important that the approximate formulae
of  Silant'ev (2002, 2007) are valid for arbitrary directed magnetic field.

Some words would be useful here
about the simplifications in our method. First of all, we
consider the optically thick plane plasma accretion disk neglecting the possible warps.
Using the Milne problem we neglect the reflection of radiation from possibly existing
central outflows (frequently this radiation lies far from observed optical wavelength bands).
In all models of magnetized accretion disks, the solutions within
the framework of the power-law dependence of magnetic field are sought inside the disk (see, for
example, Pariev et al. 2003). Physically this assumption seems fairly natural if we remember that
far from the sources the magnetic fields tend to dipole, quadrupole etc. forms, i.e.
acquire the power-law dependence. We also assume that the radial dependence of
the disk's magnetic field follows the same power-law in the range from the optical
polarized emission down to the horizon. These simplifications now are commonly
accepted, and can be considered as important assumptions of our theory.

The Milne problem in terms of vertical Thomson depth includes the possible
vertical inhomogeneities of the atmosphere, so we do not include
only possible horizontal inhomogeneities of the atmosphere. But if these inhomogeneities
are smooth (with the characteristic length of many Thomson free lengths), the corrections
should be neglected. In our paper we do not include
the true absorption effects, considering the Milne problem in the limit of conservative
atmosphere. Certainly, some our simplifications, such as the latter one, can easily be
taken into account in the proposed method. We stress that this method can be generalized
to more complex situations; in particular, it may be considered together with the
other sources of polarized radiation (polar outflows, toroidal clumpy disks, etc.).

It should be mentioned that many AGNs models postulate the existence of a dusty
geometrically thick
obscuring region ``the torus," which is placed far from the center of AGN
(see Chang et al. 2007, and many references therein). This region give
additional infrared radiation, as compared to the usual radiation of the interstellar
medium. The spectrum of linear polarization from dusty media is characterized by
Serkowski's formula (see Serkowski 1973; Martin 1989). If the spectrum of polarization
of an AGN differs strongly (as in the source NGC 4258) from Serkowski's
distribution, then the probability that the polarization comes from multiple
scattering in plasma disk increases.

Our goal in this paper is to present
the method of estimating of magnetic fields for fairly simple models. For this reason,
in particular calculations we restrict ourselves to the most popular standard disk model
of Shakura \& Sunyaev (1973).

\section{Basic equations}

We begin with the known expression for the Faraday rotation angle $\Psi$
at the Thomson optical path $\tau =\sigma_TZ$ that is frequently used below:

\[
\Psi =\frac{\omega}{2c}(n_{+}-n_{-})Z=\frac{2\pi N_ee^3B_zZ}{m^2_ec^2\omega^2}\equiv
\frac{1}{2} \delta \tau \cos{\theta}
\]
\begin{equation}
 \simeq 0.4
\left(\frac{\lambda}{1 \mu m}\right)^2 \left(\frac{B}{1 G}\right)
\tau \cos{\theta}, \label{eq1}
\end{equation}

\noindent where $n_{\pm}$ are the refractive indices for right and left circular polarized
electromagnetic waves,
 $\lambda =2\pi c/\omega$  the wavelength of the radiation and
$\theta$ the angle between the magnetic field ${\bf B}$ and line of sight
${\bf n}$ directions, $\sigma_T$ Thomson cross-section, and $N_e$ is the number density
of electrons. The first expression in Eq.~(1) is presented in many textbooks on optics.
The second expression is derived in books on plasma physics
(e.g. Kroll \& Trivelpiece 1973). This expression is usually  used in radio astronomy
(see Rohlfs \& Wilson 1996). In radiative
transfer of optical radiation, the third expression in  Eq.(1) is preferable,
because it uses well known Thomson optical depth $\tau$ (see Dolginov et al. 1995,
Gnedin \& Silant'ev 1997). According to Eq.~(1) the Faraday
dimensionless depolarization parameter $\delta$ takes a simple form:

\begin{equation}
\delta =\frac{3}{4\pi}\cdot \frac{\lambda}{r_e}\cdot
\frac{\omega_B}{\omega} \simeq 0.8\,\lambda^2(\mu m)B(G).
\label{eq2}
\end{equation}
\noindent Here $\omega =2\pi \nu=2\pi c/\lambda $ is angular frequency,
$\omega_B=eB/m_ec $ the cyclotron frequency of an electron in a magnetic field,
$r_e=e^2/m_ec^2\simeq 2.82\cdot 10^{-13}$ cm is the classic electron radius.
Parameter $\omega_B/\omega \simeq 0.93\cdot 10^{-8}\lambda (\mu m)B(G)$ is assumed
to be small. This is the condition where the simple Eq.~(1) is valid. In our
paper we have $B\approx 10^3 - 10^4 G$ near the black hole's horizon.
It this case the possible wavelength $\lambda$
can acquire values from X-rays up to infrared radiation with
$\lambda \simeq 100\mu$m.
Below we use the asymptotic formulae for the Stokes parameters, which are
valid for $\Psi\ge 1$. Therefore, the wavelength range for which our model is valid
extends from the X-ray band up to $\lambda \simeq 0.1\mu$m.
Far from the horizon
the magnetic field is less, and the depolarization parameter
$\delta\ll 1$. In this case we have to use the usual Chandrasekhar formulae
for polarization.

Silant'ev (2002) derived the asymptotical
analytical formulae for the Stokes parameters of the radiation
emitted from a magnetized, optically thick, plane-parallel
atmosphere. For the Milne problem they are

\begin{equation}
I_{\lambda} = \frac{F_{\lambda}}{2 \pi J_1}\, J(\mu), \label{eq3}
\end{equation}

\[
Q_{\lambda} =
 -\frac{F_{\lambda}}{2 \pi J_1} \,\frac{1-g}{1+g}
\,\frac{(1-\mu^2)(1-k\mu)}{(1-k\mu)^2 + (1-q)^2 \delta^2
\cos^2{\theta}},
\]

\begin{equation}
\label{eq4}
\end{equation}

\[
U_{\lambda} =
-\frac{F_{\lambda}}{2 \pi J_1}\, \frac{1-g}{1+g}
\,\frac{(1-\mu^2)(1-q)\delta\cos{\theta}}{(1-k\mu)^2 + (1-q)^2
\delta^2 \cos^2{\theta}},
\]

\noindent where $\mu = \cos{i}$ is the cosine of the angle between
the normal to atmosphere {\bf N} and the line of sight {\bf n},
$q$ the degree of true absorption ($q = \sigma_a / (\sigma_a +
\sigma_s)$),and $F_{\lambda}$ the total radiation flux. Function
$J(\mu)$ describes the angular distribution of the radiation
emerging from a disk. This function, as well as the numerical
parameters $g$, $k$, and $J_1$ were tabulated by Silant'ev (2002).
For electron conservative atmosphere ($q=0$), the
values of these parameters are $k = 0$, $g = 0.83255$, and $J_1=1.19402$.
The Stokes parameters $Q$ and $U$ are given in the reference frame with
the X-axis lying on the plane $({\bf nN})$ (see Fig.~1).

Formulae ~(3--4) for polarization consider the last scattering
of radiation  before escaping from a semi-infinite magnetized atmosphere.
For a high value of parameter $\delta $, the contribution of the secondary
scattered photons is small because of the large Faraday depolarization. Even
in the absence of a magnetic field, the main contribution to the polarization
of emitted radiation comes from the last scattered photons. For this reason,
Eqs.~(3--4) at the absence of magnetic field practically represent
the classical Chandrasekhar-Sobolev polarization in the Milne problem
(see, for example, Chandrasekhar 1950).

As far as the intensity
of radiation $I(\mu)$ is concerned, one can remember (see Chandrasekhar 1950) that
the polarization weakly influences the intensity. For Milne's problem without
the true absorption ($q=0$), we have $I(0)\sim 3.06$, whereas the separate
transfer equation with the Rayleigh phase function gives $I(0)\sim 3.02$. For high
values of $\delta $, the terms with Stokes parameters $Q$ and $U$ in the full system
of transfer equations for parameters $I, Q$ and $U$ become very small $\sim 1/\delta$,
and they are negligible in the equation for intensity $I$. As a result,
the radiation intensity obeys the separate transfer
equation with the Rayleigh phase function (see, Silant'ev 1994 for more detail).
Expression (3) presents the solution to this equation. For large $\delta$
the main contribution to polarization comes from to intensity term. Formulae (4) were
obtained in this way.

Equations~(2--4) allow us to derive the following approximate
expressions for polarization degree $p({\bf B},{\bf n})$ and
the position angle $\chi $ of radiation for an accreting magnetized disk:

\[
p({\bf B},{\bf n}) \simeq
\frac{p(0,\mu)}{\sqrt{(1-k\mu)^2+(1-q)^2\delta^2\cos^2\theta }},
\]
\begin{equation}
\tan{2\chi }=\frac{U_{\lambda}}{Q_{\lambda}}\simeq
\frac{(1-q)\delta}{1-k\mu} \cos\theta, \label{eq5}
\end{equation}

\noindent where $B \cos\theta = {\bf B n}$.
Now we consider the case of dominant nonabsorbing electron scattering in accretion disks,
i.e. $q=0$ and $k=0$.
Photons escape the optically thick disk basically from the surface
layer with $\tau\approx 1$. If the Faraday rotation angle $\Psi$
corresponding to this optical length becomes larger than unity,
then the emerging radiation will be depolarized as a result of
summarizing the radiation fluxes with the very different angles
of Faraday rotation. Only for directions that are nearly
perpendicular to the magnetic field are the Faraday rotation angles
too small and depolarization does not occur. Certainly, the
diffusion of radiation in the inner parts of a disk depolarizes
the light, even in the absence of a magnetic field, because of multiple
scattering of photons.

The existence of a magnetic field, hence Faraday rotation,
only increases the depolarization process. It means that the
polarization of outgoing radiation acquires a peak-like angular
dependence with its maximum for perpendicular propagation. The
sharpness of the peak increases with increasing magnetic field
magnitude. The main region of allowed angles appears to be $\sim
1/\delta $.

Another very important feature of the polarized
radiation is the wavelength dependence of polarization degree $p$
and position angle $\chi$ which is very different from
the case of classical electron
scattering. This effect is briefly considered in Sect. 5.
We consider the case where the Thomson
cross-section does not depend on the radiation wavelength.

\subsection{Integral polarization from the accretion disk}

The axially symmetric accretion disks frequently are observed as
a whole. The observed integral Stokes parameters $\langle Q\rangle $
and $\langle U\rangle $ are described by the azimuthally averaged
formulae~(3) and (4). To derive these expressions we introduce the
following notions:

\begin{equation}
\delta \cos\theta=\delta_{\|}\cos i+\delta_{\bot}\sin i
\cos(\varphi+\varphi_*)\equiv a+b\cos\Phi, \label{eq6}
\end{equation}
\noindent where $\varphi $ is the azimuthal angle
of radius-vector ${\bf r}$ (the azimuthal angle of line of sight
${\bf n}$ is taken zero), $\varphi_*$ the angle between perpendicular
magnetic field projection ${\bf B}_{\bot}=(B_{\rho}, B_{\varphi})$, and
the projection ${\bf B}_{\rho}$($\cos\varphi_*=B_{\rho}/B_{\bot}$)
(see Fig.~1).
The value $B_{\|}\equiv B_z$ is the magnetic field projection along the
normal ${\bf N}$ to the accretion disk plane. According to Eqs.~(1) and (6)
we have
\[
a=\delta_{\|}\mu ,\,\,\,\, \delta_{\|}=0.8\lambda^2(\mu m)B_{\|}(G),
\]
\begin{equation}
b=\delta_{\bot}\sqrt{1-\mu^2},\,\,\,\,
\delta_{\bot}=0.8\lambda^2(\mu m)B_{\bot}(G). \label{eq7}
\end{equation}
\noindent Using the axial symmetry of an accretion disk, we obtain
(remember that $ U(0,\mu)\equiv 0$ for Milne problem without magnetic
field):

\[
\langle Q\rangle =Q(0,\mu)\times
\]
\[
\frac{2}{\pi}\int_0^{\pi/2}d\Phi
\,\frac{1+a^2+b^2\cos^2\Phi}
{(1+a^2+b^2\cos^2\Phi)^2-(2ab\cos\Phi)^2},
\]

\[
\langle U\rangle =a\,Q(0,\mu )\times
\]
\begin{equation}
\frac{2}{\pi}\int_0^{\pi/2}d\Phi \,\frac{1+a^2-b^2\cos^2\Phi}
{(1+a^2+b^2\cos^2\Phi )^2-(2ab\cos\Phi )^2}. \label{eq8}
\end{equation}
\noindent The observed degree of the light polarization and the position
angle are derived from parameters (8) in the usual way. For particular
cases of pure normal ($\delta_{\bot}=0$) and pure perpendicular
($\delta_{\|}=0$) magnetic fields, expression (8) can be derived
analytically. For the first case we have

\begin{equation}
p({\bf B},{\bf n})=\frac{p(0,\mu)}{\sqrt{1+\delta^2_{\|}\mu^2}},
\,\,\,\,\,\, \tan2\chi=\delta_{\|}\,\mu \,. \label{eq9}
\end{equation}
\noindent For perpendicular magnetic field are there the formulae
\begin{equation}
p({\bf B},{\bf n})=\frac{p(0,\mu)}{\sqrt{1+\delta^2_{\bot}
(1-\mu^2)}},\,\,\,\,\,\, \chi\equiv 0. \label{eq10}
\end{equation}
\noindent For a pure perpendicular magnetic field, the position angle $\chi=0$
comes from the axial symmetry of the problem, so the electric wave
oscillations in this case occur parallel to the surface of an accretion disk.
The case $\chi\neq 0$ can be realized if some $B_z$ component exists.
It is interesting that the cases  $(B_{\rho}\neq 0, B_{\phi}=0)$ and
$(B_{\rho}=0, B_{\phi}\neq 0)$ give rise to the same formula (10). This is the
consequence of the averaging procedure.

It is seen from Eqs.~(8) that the
relative degree of polarization $p({\bf B},{\bf n})/p(0,\mu)$ and position angle $\chi$
only depend on
dimensionless parameters $a$ and $b$, which are the functions of wavelength $\lambda$,
magnetic fields $B_{\|}$ or $B_{\bot}$, and inclination angle $i$ ($\mu =\cos i$).
First we discuss the behavior of the relative degree of polarization and
position angle on these two parameters. Remember that we consider conservative atmosphere with
$q=0$, hence $k=0$. It is interesting to investigate how the
polarization changes if we include the perpendicular magnetic field $B_{\bot}$
in the existing parallel magnetic field $B_{\|}$.

\begin{figure*}[t!]
\centering\leavevmode \epsfxsize=0.85\textwidth
\epsfbox{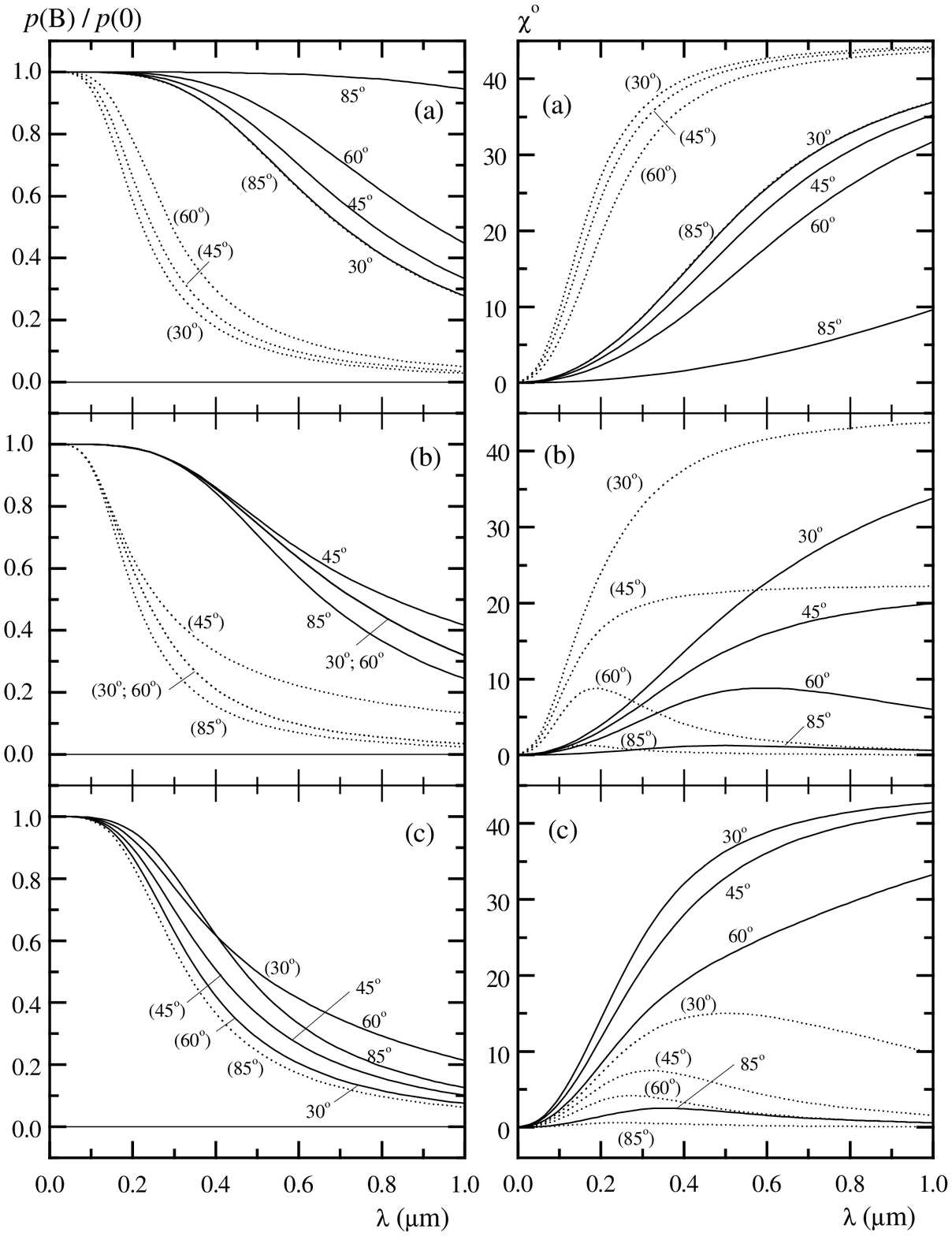} \caption{\small Spectra of relative
polarization degree $p({\bf B},{\bf n})/p(0,\mu)$ and position
angle $\chi $. The numbers denote the values of inclination angle
i. The numbers in brackets refer to dotted curves. The a) figures
demonstrate spectra for the pure parallel magnetic field with
$B_{\|}=5$ G (bold curves) and 50\,G (dotted curves). The b)
figures concern the case $B_{\|}=B_{\bot}=5$ G (bold curves) and
50\,G (dotted curves). The c) figures present spectra for
$B_{\|}=20$ G and $B_{\bot}=10$ G (bold curves), and for
$B_{\|}=10$ G and $B_{\bot}=20$ G (dotted curves).} \label{fig2}
\end{figure*}

The average process takes the disk regions with the very
different angles $\theta$ into account between the magnetic field ${\bf B}$ and the
line of sight ${\bf n}$, and the integral polarization and position
angle can acquire very different values. The numerical calculations
show that position angle $\chi $ not only depends on parameter
$a=\delta_{\|}\mu $ but also on the $b$ - parameter
($b=\delta_{\bot}\sqrt{1-\mu^2}$). The existence of perpendicular magnetic
field ${\bf B}_{\bot}$ diminishes the value of $\chi $ compared to the case
of pure parallel magnetic field. This decrease is especially large if $b>a$
and the parameters $a$ and $b$ are close to unity. For $a\gg 1$ and
$b<a$ the decrease of $\chi $ is small and practically $\chi\simeq 45^{\circ}$.
But for $b>a\gg 1$, the position angle $\chi\to 0$. The
special case is $a=b\gg 1$. In this case the position angle $\chi$
rapidly decreases from the limiting value $\sim 45^{\circ}$ for
$a-b\simeq 1\div 5$  to value $22.5^{\circ}$ at $b=a$, and then tends to zero
for $b-a\simeq 1\div 5$, so for large $a$ the intermediate values of $\chi $
can  only occur in a rather narrow interval $b-a\simeq 1\div 2$, i.e. at
$a\simeq b$.

The degree of linear polarization $p$ depends on parameters $a$ and $b$
in a more complex form.
For $a\le 1$ the addition of the perpendicular magnetic
field (parameter $b$) lowers the integral polarization. For $a>1$
there is the region of $b$ ($b<a$) where the polarization increases
compared with the case of pure parallel magnetic field. The maximum
polarization occurs at $b\simeq a$, and then the polarization decreases with
the increase in parameter $b$. The increase in polarization at $b=a$ can
be rather large. As a result, for value $a=b=5,10,20$, and $50$ the relative polarization
increase, as compared to purely parallel magnetic field, is equal to 160\%, 224\%,
317\%, and 504\%, respectively.
It seems this effect stems some ``resonant" regions in
an accretion disk where the Faraday rotation from parallel magnetic field
is balanced by opposite rotation from a perpendicular magnetic field. Of course,
the magnitude of polarization decreases with the increase in $a$ and $b$.
The numerical calculations demonstrate that the relative polarization degree
$p({\bf B},{\bf n})/p(0,\mu)$ is a symmetric function of parameters $a$
and $b$. The position angle $\chi $ does not possess this symmetry.

Now we shortly discuss the wavelength dependence of polarization degree $p(\lambda)$
and $\chi(\lambda)$, which follows from general formulae (8). More detailed discussion is
presented in Sect. 5. For high values
of parameters $a$ and (or) $b$ the spectra diminish $\sim 1/\lambda^2$.
But for the case  $a=b$ mentioned above, the spectra diminish as $\sim 1/\lambda$.
Thus, the ``resonant" effect also changes the asymptotic behavior of spectra.
Note once more that this effect disappears beyond the interval $|a-b|\approx 1-5$.
The spectra $\chi(\lambda)$ depend strongly on the relative value of
the perpendicular magnetic field $B_{\bot}$ as compared to vertical
component $B_{\|}$.
If parameter $a \gg 1$ and $a \gg b$ the position angle $\chi(\lambda)\to 45^{\circ}$;
i.e., it becomes independent of wavelength. For the ``resonant" case $a=b$, this limiting value
is equal to $22.5^{\circ}$.

The characteristic spectra of polarization and position
angle as a function of inclination angle $i$ and magnetic fields $B_{\|}$ and
$B_{\bot}$ are presented in Fig.~2 ($\mu=\cos i$). The mentioned case $a=b$ corresponds
to $B_{\|}\mu =B_{\bot}\sqrt{1-\mu^2}$; i.e., at $B_{\|}=B_{\bot}$ it exists at
$i=45^{\circ}$. For every $\lambda$ there exists its own value
$a(\lambda)=b(\lambda)$, so, for $\lambda=1\mu$m and $B_{\|}=B_{\bot}=5$G, this value is
$a=b=2.828$.
In this figure we present the relative polarization degree
$p({\bf B},{\bf n})/p(0,\mu)$ for
the inclination angles $i=85^{\circ}, 60^{\circ}, 45^{\circ}$, and $30^{\circ}$.
The values of the polarization degree
$p(0,\mu)$ for these angles are equal to $7.80\%, 2.25\%, 1.08\%$,
and $0.43\%$, respectively (see Chandrasekhar 1950). The presented
spectra can help readers  recognize general tendencies of polarization
as a function of the basic system parameters.

We see that the
higher magnetic field, the larger depolarization. For
$i\approx 90^{\circ}$ and pure parallel magnetic field (${\bf B}_{\bot}=0$),
the polarization degree tends to the Thomson value of polarization $p(0,\mu)$.
This is quite natural because in these cases the magnetic field is
practically perpendicular to the line of sight ${\bf n}$, and the Faraday
rotation is low. The position angle $\chi$ is more sensitive to Faraday
rotation, and tends to Thomson value ($\chi =0$) slower than the polarization
degree tends to the Thomson polarization. It is interesting that for this case
the relative polarization degrees $p({\bf B},{\bf n})/p(0,\mu)$
and position angles $\chi$ practically coincide
for inclination angles $i=85^{\circ}$ and $i=30^{\circ}$, if the
magnetic fields differ 10 times (for example, $B_{\|}=5$ and 50, or
$B_{\|}=10$ and 100, etc.). This happens because the corresponding values
of $\mu =\cos i $, 0.08715 and 0.86602, differ approximately 10 times.

Comparing formulae (9) with (10), we find that the relative polarization
degrees for perpendicular magnetic field ($B_{\|}=0)$ can be taken from
the results drawn in Fig.~2a, if one uses the substitution
$i\to (90^{\circ}-i)$ there. The case $B_{\bot}=5$ G  and $i=60^{\circ}$
therefore coincides with the case $B_{\|}=5$ G and $i=30^{\circ}$. Of course, the
position angle $\chi=0$ for perpendicular magnetic field.

For the case $B_{\|}=B_{\bot}$ (see Fig.~2b) the most depolarization
occurs at $i\simeq 90^{\circ}$, in contrast to the pure parallel
magnetic field. It is interesting that the
relative degrees of polarization are the same for the inclinations
$i$ and $90^{\circ}-i$. But the position angles are different in these
cases. The relative polarization degree is higher for $i=45^{\circ}$
than for $i=30^{\circ}$, i.e. the  change in this value is not monotonic.
The case $i=45^{\circ}$ corresponds to equality $a=b\gg 1$. As
mentioned above, in this case $\chi \to 22.5^{\circ}$. The righthand side
of Fig.~2b confirms this.

Figure 2c presents the spectra of the relative polarization degree and position
angle for $B_{\|}=20$ G and $B_{\bot}=10$ G (the curves denoted by the usual
numbers) and the opposite case $B_{\|}=10$~G and $B_{\bot}=20$~G (the curves
denoted by numbers in brackets). The symmetry of
relative polarization degree as a function of parameters $a$ and $b$ gives rise
to the coincidence of spectra in the first case with those in the second case if
the angle $i\to (90^{\circ}-i)$, so the relative polarization degree spectra
at $30^{\circ}, 45^{\circ},60^{\circ}$ and $85^{\circ}$ coincide with
the spectra denoted as $(60^{\circ}), (45^{\circ}), (30^{\circ})$ and
$(5^{\circ})$, respectively. But the position-angle spectra are different for
these two cases.

The spectra, presented in Fig.~2, demonstrate a large variability of values
and forms for polarization degrees and position angles in the integral
radiation escaping from the magnetized accretion disks.

\subsection{Some results from accretion disk models}

Models of a magnetic accretion disk with externally imposed, large-
scale vertical magnetic field and anomalous magnetic field
diffusion due to enhanced turbulent diffusion have been considered
by Campbell (2000), Ogilvie \& Livio (2001), and Pariev et al. (2003).

We calculate the value of Faraday depolarization parameter
$\delta$ for the model of an accreting disk suggesting the
power-law radial dependence of the magnetic field:

\begin{equation}
B(r) = B_H(R_H/r)^n, \label{eq11}
\end{equation}

\noindent where $B(r)$ is the magnetic field inside an
accretion disk, and $B_H$ the magnetic field strength at the event
horizon radius of a SMBH. The radius of the black hole horizon is
(Novikov \& Thorne 1973)

\begin{equation}
R_H = \frac{G M_{BH}}{c^2}\left(1 + \sqrt{1 - (a/M_{BH})^2}\right)
\label{eq12},
\end{equation}

\noindent where $a/M_{BH}$ is the spin of a Kerr black hole.

Pariev et al. (2003) have developed the detailed description of magnetized
accretion disks with the different values for parameter $n$ (denoted as $\delta$
in their paper). The radial effective temperature dependence $T_e \sim r^{-3/4}$ in
their model is the same as in the Shakura-Sunyaev model. They find that, in the case
of equipartition of magnetic pressure with radiation or thermal pressures, their
results are close to Shakura-Sunyaev model with the viscosity parameter $\alpha=1$.
As a most
physically significant model, they investigate the case $n=5/4$ in more detail.
Parameter $n$ must be greater than unity if we consider the disks
with the diminishing gas density far from the black hole. It should also be noted
that dependence (11) with $n=5/4$ takes place at the equipartition of magnetic
energy with thermal and gravitational ones in the spherical accretion (see, for
example, Melia 1992). Following to Pariev et al. (2003), we use mostly
the case $n=5/4$.

The central problem is to derive the characteristic scale
$R_{\lambda}$ that corresponds to the effective wavelength of
polarimetric observations. At first glance, we can estimate the
radius $R_{\lambda}$, suggesting that $\lambda = \lambda_e$
corresponds to the value of rest-frame wavelength of the black body
spectrum maximum,

\begin{equation}
\lambda_e = \frac{0.29}{T_e(r)},\,\,\,\,\,\, \lambda_0 =
\lambda_e(1+z), \label{eq13}
\end{equation}

\noindent where $\lambda_0$ is the wavelength in the observer
system, and $z$ the cosmological redshift.

In a standard thin disk model (Shakura \&
Sunyaev 1973), there are a black body radiation with an effective
temperature profile of $T_e = T_H(r/R_H)^{-3/4}$ and the scale
length $R_{\lambda}$ defined by the point in the disk where the disk temperature
matches the rest-frame wavelength of the monitoring band.

There is the series of papers where the semi-empirical method of
determining of the accretion disk scale $R_{\lambda}$ has been developed (see
Kochanek et al. 2006; Poindexter et al. 2008; Morgan et al. 2007, 2008).
The authors used microlensing variability observed for
gravitationally lensed quasars to find the accretion disk size and
the observed (or rest-frame) wavelength relation. It is very
important that the scaling appeared to be consistent with what is
expected from the thin accretion disk theory of Shakura \& Sunyaev (1973).

This allows us to have the following size scaling (Poindexter et al.
2008):

\begin{equation}
R_{\lambda}(cm) = 0.97\times 10^{10} \left(\frac{\lambda_e}{\mu m
}\right)^{4/3} \left(\frac{M_{BH}}{M_{\odot}}\right)^{2/3}
\left(\frac{\eta}{\varepsilon}\right)^{1/3}. \label{eq14}
\end{equation}

\noindent Here $R_{\lambda}$ is the distance in the accretion
disk, which corresponds to $\lambda_e$; $\eta = L_{bol}/L_{Edd}$,
$L_{Edd} = 1.3\times 10^{38} (M_{BH}/M_{\odot})$ erg s$^{-1}$ as the
Eddington luminosity, and $\varepsilon$ as the rest-mass radiation
conversion efficiency. Note that $R_{\lambda}$ does not depend on
viscosity parameter $\alpha$.
We use the commonly accepted relation
between the bolometric luminosity $L_{bol}$ of the accretion disk
and the accretion rate $\dot{M}$:

\begin{equation}
L_{bol} = \varepsilon \dot{M} c^2. \label{eq15}
\end{equation}

Theoretical calculations of parameter $\varepsilon$ depend on the details
of the accepted model. Usually one presents the dependence of $\varepsilon$ on
the spin parameter $a/M_{BH}$ for every particular model. Naturally, this
dependence is different for different models. Krolik (2007) present the
comparison of $\varepsilon$-values from Novikov-Thorne model (Novikov\&Thorne 1973)
and the model that take the jet luminosity into account. For spin parameter
$a/M_{BH}=0.5$, the first model gives $\varepsilon=0.081$, whereas the second model gives
$\varepsilon=0.0063$. In some papers
one considers $\varepsilon$ as an independent parameter in the model calculations.
For example, Pariev \& Colgate (2007) accept $\varepsilon =0.1, \eta=0.1$ and
viscosity parameter $\alpha=0.01$.

The strong gravitational field near the black
hole influences the Stokes parameters of outgoing radiation when they propagate
to an observer. The detailed calculations of this effect have been done by
Connors et al. (1980), Karas et al. (2004), and Dovciak et al. (2004). Usually
these effects are  important when describing  X-ray emission from the vicinity
of the black hole. The optical radiation arises far from this place, and we neglect
these gravitational corrections.

\section{Magnetic field strength of NGC 4258}

NGC 4258 is a low-luminosity Seyfert II galaxy at the distance of
about 7.2 Mpc, which harbors water masers (Modjaz et al.
2005). This object is usually considered as a very good
laboratory for successfully measuring the magnetic field in accretion disk
even very close to the central black hole.
The spatial velocity distribution of water mega-maser sources in
NGC 4258 on scales of 0.14-0.28 pc indicates a thin Keplerian disk
rotating around a black hole with a mass $M_{BH} = 3.9\times
10^7 M_{\odot}$ (Herrnstein et al. 1999). The
accretion disk has an almost edge-on orientation with the
radiation axis and its inclination angle is $i = 83^{\circ}\pm
4^{\circ}$ (Pringle et al. 1999). A pc-scale jet closely
aligns in projection on the sky with the rotation axis (Herrnstein
et al. 1997).

Modjaz et al. (2005) present an analysis of polarimetric
observations at 22 GHz of the water vapor masers in NGC 4258
obtained with the VLA and the GBT. They do not detect any
circular polarization in the spectrum indicative of Zeeman-induced
splitting of the maser lines of water, and obtained only an upper
limit on the magnetic field strengths. They obtained the
$1-\sigma$ upper limit value of the toroidal component of the
magnetic field at a radius of 0.2 pc the value of 90 mG and
determined a $1-\sigma$ upper limit of 30 mG on the radial
component at a radius of 0.14 pc. They also find from their observations
of magnetic field limits that the geometrically thin disk model and the
jet-disk model are better candidates for accounting for the extremely
low-luminosity of NGC 4258.

More recently, Reynolds et al. (2008) have shown from analysis of
SUZAKU and XMM-Newton observational data that the observed iron lines
originate in the surface layers of an warped accretion disk at
the distance $10^3-10^4R_H$ from the black hole.
In contrast to the majority of Seifert 2 galaxies, there was no
indication of a Compton-thick obscuring torus. The weak iron line and
the lack of a reflection point to circumnuclear environment that is
remarkably clean of cold gas. They note that such a circumnuclear environment
is only found in two AGNs - NGC 4258 and M81 that contrast to the
majority of Seifert 2 galaxies.

This picture coincides with one by Herrnstein et al.(2005)
pointed out earlier. According to them the observing intrinsic
absorption in the X-ray spectrum can arise in
the outer layers of the warped geometrically-thin accretion disk at the distance
$\sim 29$ pc from the black hole, where the molecular-to-atomic transition occurs.

This picture allowed us to use our simple model of arising of optical
polarized radiation without taking the warp contribution into account.
Of course, the power-law dependence of the magnetic field is an important
assumption in our derivation. This assumption
is now commonly accepted (see Pariev et al. 2003).

The detected of polarized continuum and line emission from the
nucleus of NGC 4258 was by Wilkes et al.
(1995). After that, Barth et al. (1999)
obtained spectropolarimetric observations of the NGC 4258 nucleus
at the Keck II telescope. The observations were obtained on 1997
April 10 UT at the Keck II telescope with the LRIS
spectropolarimeter. The results of these observations are
presented in the Table 1 of the paper by Barth et al.
(1999). For the continuum polarization they obtained the
following results:

\[
p(\lambda\lambda\, 4000\div 4800\, \dot{A}) = 0.38\pm
0.03\%,\,\, \chi = 12^{\circ}\pm 2^{\circ}
\]
\[
p(\lambda\lambda\, 5100\div 6100\, \dot{A}) = 0.35\pm
0.01\%,\,\, \chi = 7^{\circ}\pm 1^{\circ}
\]
\begin{equation}
p(\lambda\lambda\, 7500\div 8500\, \dot{A}) = 0.29\pm 0.02\%,\,\,
\chi = 8^{\circ}\pm 2^{\circ} \label{eq16}
\end{equation}

\noindent We see that the polarization is weakly increasing to the short
wavelength range, but the value of a position angle is practically
constant. The position angles $\chi $ are the angles
between wave electric field oscillations and the surface of the
accretion disk.

\subsection{Estimates of magnetic field in the model of
nonturbulent accretion disk}

For the inclination angle $i = 83^{\circ}$, the
expected polarization should have the value $p(0,\mu) = 6.9 \%
\, (\mu = \cos{i} = 0.122)$. From Eqs.~(9) and (10)
for degree of polarization
we find that possible parameters $a=\delta_{\|}\mu$ and
$b=\delta_{\bot}\sqrt{1-\mu^2}$ are near the value 20. The
position angle $\chi $ for this value of $a$ is equal to $43.6^{\circ}$,
which is far from observing values (see Eq.~(21)). From the discussion
in Sect.~2.1 we know that for high values of parameters $a$ and $b$,
the possibility of small position angles exists if $a\simeq b$.

The exact formulae (8) give us the values
$a=122$ and $b=122.9$, which correspond to observed values of
polarization degree 0.38\% and $\chi=12^{\circ}$ at $\lambda =0.44\mu m$.
These values correspond to $\delta_{\|}=1000$ (or $B_{\|}=6400$ G ) and
$\delta_{\bot}=124$ (or $B_{\bot}=800 $ G). The analogous values of
parameters for the polarization degree 0.35\% and $\chi =7^{\circ}$ at
$\lambda =0.56 \mu m$ are $\delta_{\|}=762.3$ (or $B_{\|}=3037$ G) and
$\delta_{\bot}=95.5$ (or $B_{\bot}=380.5$ G). For the case
$\lambda =0.8 \mu m$ ($p=0.29\%, \chi=8^{\circ}$), the formulae (8)
give $B_{\|}=2401.4$ G and $B_{\bot}=298.3$ G. We see that in all cases
the normal magnetic field $B_{\|}$ is greater than $B_{\bot}$. Because
the effective temperature $T_e$ decreases with the increase in the
distance from the inner radius of an accretion disk, we conclude that
magnetic field also decreases with the growing distance.

What seems unsatisfactory in these results is their
sensitivity to small variations in parameters $a$ and $b$. If these
parameters change their values ($\Delta(b-a)\simeq 1$) slightly the
solution is impossible. For magnetic fields it means that the values
have not to change its values greater than 10 G. This is very
improbable for real situations in accretion disks. For this reason
we have to seek a more satisfactory model where this sensitivity does not
occur. Such a model really exists. It takes into account that the magnetic
field can be turbulent.

\subsection{Estimates of magnetic field in the model of turbulent
accretion disk}

According to Silant'ev (2005, 2007) the chaotic
component ${\bf B'}$ of the magnetic field (${\bf B}={\bf B}_0+{\bf B'}$),
where ${\bf B}_0$ is a regular part of the magnetic field,
gives rise to additional extinction of the intensity of linearly polarized
waves (parameters $Q$ and $U$) due to small scale chaotic Faraday rotations.
The Gaussian distribution of turbulent velocities was assumed. Mathematically,
this effect is analogous to the known problem of diffusion of scalar impurity
in a stochastic velocity field (see, for example, van Kampen 1981). In our case,
the Faraday rotation angle replaces the role of impurity.
The main part of the effective cross-section $\sigma_{*}\equiv \sigma_T\,C$
corresponding to this additional extinction, takes a very simple form:

\begin{equation}
C=f_B\,\tau_1\langle \delta'^{2}\rangle\simeq
0.64f_B\,\tau_1\lambda^4(\mu m) \langle B'^{2}(G)\rangle/3.
\label{eq17}
\end{equation}
\noindent Here, $\tau_1$ is mean Thomson optical radius of turbulent eddies,
$f_B\simeq 1$ is a constant characterizing a particular form of two-point
turbulent velocity correlations (for estimations we take $f_B=1$), and
 $\delta'$ and ${\bf B'}$ are fluctuating parts
of Faraday depolarization parameter $\delta$ and magnetic field ${\bf B}$,
respectively. Clearly, the additional extinction should be proportional
to the mean square of fluctuations of physical parameter $\delta$. Because
$\delta\sim \lambda^2$, the dimensionless parameter $C\sim \lambda^4$.

The asymptotic formulae taking this effect into account have the same form
as formulae (3-5) with the substitution $(1-k\mu)\to (1+C)$.
(Remember that we consider the conservative atmosphere with $q=k=0$.)
In particular, the formulae (8) acquire the form

\[
\langle Q\rangle =(1+C)\,Q(0,\mu)\times
\]
\[
\frac{2}{\pi}\int_{0}^{\pi/2}d\Phi
\frac{(1+C)^2+a^2+b^2\cos^2\Phi}{[(1+C)^2+a^2+b^2\cos^2\Phi]^2
-(2ab\cos\Phi)^2}
\]
\begin{equation}
\langle U\rangle =a\,Q(0,\mu)\times \label{eq18}
\end{equation}
\[
\frac{2}{\pi}\int_{0}^{\pi/2}d\Phi
\frac{(1+C)^2+a^2-b^2\cos^2\Phi}{[(1+C)^2+a^2+b^2\cos^2\Phi]^2
-(2ab\cos\Phi)^2}.
\]

\noindent Here parameters $a$ and $b$ are presented by formulae (7), where $B_{\|}$
and $B_{\bot}$ denote the regular (mean) values of corresponding magnetic fields, i. e.
$B_{0\|}$ and $B_{0\bot}$.
The numerical calculations show that the relative polarization degree
$p({\bf B},{\bf n})/p(0,\mu)$ is also the symmetric function of
parameters $a$ and $b$, as for the nonturbulent atmosphere.

The parameter $C$ is fairly large for our case, $C\approx 10\div 20$. In this
case the term $(2ab\cos\Phi)^2$ in denominators of formulae (18) can be
neglected and we obtain the following analytical expressions:

\[
\tan2\chi \cong \frac{a}{(1+C)(1+d)},\,\,\,\, d=\frac{b^2}{(1+C)^2+a^2},
\]
\begin{equation}
p\cong \frac{p(0,\mu)}{[(1+C)^2+a^2]\sqrt{(1+d)}}
\sqrt{(1+C)^2+\frac{a^2}{(1+d)^2}}. \label{eq19}
\end{equation}
\noindent These expressions are fairly exact up to $a,b<C$. If parameters $a$ or $b$
are equal to zero, formulae (19) are exact. In particular,
for the cases of pure normal regular magnetic field ${\bf B}_{0\bot}=0$ ($b=0$),
and pure perpendicular regular field ${\bf B}_{0\|}=0$ ($a=0$), they transform to the
following exact formulae:

\[
p({\bf B},{\bf n})=\frac{p(0,\mu)}{\sqrt{(1+C)^2+\delta^2_{\|}\mu^2}},
\,\,\,\,\, \tan2\chi=\frac{\delta_{\|}\mu}{1+C},
\]
\begin{equation}
\label{eq20}
\end{equation}
\[
p({\bf B},{\bf n})=\frac{p(0,\mu)}{\sqrt{(1+C)^2+\delta^2_{\bot}(1-\mu^2)}},
\,\,\,\, \chi\equiv 0.
\]
\noindent In the limit case of nonturbulent magnetic field ($C=0$), they coincide
with the formulae (9) and (10). If the chaotic magnetic field $B'$ is fairly
high ($\tau_1 \lambda^2(\mu m)\langle B'^2\rangle \ge 4B_{0\|,0\bot}$, see Silant'ev 2007),
then the depolarization of radiation is due to additional extinction (17):
$p(B',{\bf n})=p(0,\mu)/C\,\sim 1/\lambda^4\langle B'^2\rangle$.
If we suppose that $\langle B'^2\rangle=Const $,
then $p(\lambda)_{turb}\sim \lambda^{-4}$.

The existence of
new parameter $C$ describing the level of magnetic field fluctuations
makes the estimation of mean values of $B_{\|}$ and $B_{\bot}$ difficult.
In our case this problem is simpler because the level of magnetic
fluctuations (parameter $C$) changes slowly with the variations in parameters
$a$ and $b$. Indeed, for the pure normal magnetic field ($b=0$) we found
$a=7.37, \,C=15.5$ for $\lambda =0.44\mu m$; $\,a=4.77, \,C=18.13$ for
$\lambda=0.56 \mu m$, and $a=6.56, \,C=21.88$ at $\lambda=0.8\mu m$.
In the limiting case of high values of polarization degree ($a=b$), these
values are: $a=b=8, \,C=14$; $a=b=5, \,C=17.5$, and $a=b=7, \,C=21$,
respectively; i.e., parameter $C$ is practically the same for these two cases.

For this reason we calculate the $a$ and $b$ parameters using the mean
values for parameter $C$; i.e., $C=14.75, 17.81$ and $21.44$, respectively.
This gives the values $a=8, \,b=6$ ($B_{\|}=423 \,G,\,
B_{\bot}=39$ \,G) for $\lambda=0.44\mu m$; for $\lambda=0.56\mu m$ -- $a=5,
\,b=3.5$ ($B_{\|}=163 \,G,
B_{\bot}=14 G$); and for $\lambda=0.8 \mu m$ -- $a=7, \,b=5.5$
($B_{\|}=112 \,G,
B_{\bot}=10.8\,G$). These values of magnetic fields are lowere
than those for the case of nonturbulent accretion disk. But important is that
they were derived without  restriction $a\simeq b$.

Let us estimate the
level of magnetic fluctuations taking the mean values for parameter $C$.
We also assume that $f_B=1$ and $\tau_1=0.1$. We do not know the real distribution
of turbulent eddies in a turbulent accretion disk. The general picture of turbulence
consists of cascade of eddies with different dimensions. The small
eddies with $\tau_1\ll 1$ includes a value $propto$ a parameter $C$. The large
eddies do not allow us to describe the considered effect in the range of radiative
transfer equations. It seems that our value $\tau_1\simeq 0.1$ is fairly natural for
describing the turbulent effects in magnetized plasma. If we increase $\tau_1$
to k-times,
then the level of magnetic fluctuations $\langle B'^2\rangle$ decreases to k-times.
The real value of parameter $\tau_1$ can therefore be estimated by independent estimation of
magnetic fluctuations.
After that we obtain the values
$B'\approx 136 \,G, \,92 \,G$, and $49 \,G$ for the mean square root values of magnetic
fluctuations at places where the thermal radiation
has a maximum for $\lambda =0.44\mu m, 0.56\mu m$, and $0.8\mu m$,
respectively. These values of fluctuations are equal to 31\%, 56\%, and 44\% of
the mean magnetic fields for mentioned wavelengths.

\subsection{Estimates of the magnetic field at the horizon
of the black hole in NGC 4258}

We now proceed to the estimation of the magnetic field strength in
NGC 4258 using the maser polarimetric data.
The magnetic field structures in accretion disks are difficult
to observe and remain poorly known. If the disk is penetrated by a
dipole field of the central object or by a global field of the
surrounding interstellar medium, there may be a net vertical flux.
Sano et al. (2004) consider the models of an accretion
disk with a uniform magnetic field. The stress forces in accretion
disks may be proportional to $B_z$ (Hawley et al. 1995)
or to $B_z^2$ (Turner et al. 2003).

Zhang \& Dai (2008) have studied the effect of a
global magnetic field on viscously rotating and vertically
integrated accretion disks around compact objects using a
self-similar treatment. They show that the strong magnetic
field in the vertical direction prevents the disk from being
accreted and decreases the effect of the gas pressure.

On the other hand, K\"onigl (1989) and Cao (1997)
underline that the inclination of the field lines at the surface
of the disk plays a crucial role in the magnetically
driven outflow. They show that, for the nonrelativistic case, a
centrifugally driven outflow of matter from the disk is only
possible if the poloidal component of the magnetic field makes an
angle less than a critical $60^{\circ}$ with the disk surface.

Now let us estimate the value of $R_{\lambda}$ for the model of a
standard accretion disk using the observational data of NGC 4258.
The spatial structure of standard accretion disk have been
calculated by Poindexter et al. (2008). The size
scaling is determined by Eq.~(14). The basic physical
parameters of the central nucleus of NGC 4258 are $M_{BH} =
3.9\cdot 10^7 M_{\odot}$ and $\eta = L_{bol}/L_{Edd} =
10^{-2.27}= 0.0054$ (Satyapal et al. 2005). We use these
estimations below in numerical calculations.

The estimates of the scaleradius $R_{\lambda}$ from
Eq.~(14) give the following results:

\[
R_{\lambda} = 1.38\cdot 10^{14}
\left(\frac{0.1}{\varepsilon}\right)^{1/3}\, cm,\,\,
\lambda_{rest} = 0.44\, \mu m,
\]
\[
R_{\lambda} = 1.9\cdot 10^{14}
\left(\frac{0.1}{\varepsilon}\right)^{1/3}\, cm,\,\,
\lambda_{rest} = 0.56\, \mu m,
\]
\begin{equation}
R_{\lambda} = 3.06\cdot 10^{14}
\left(\frac{0.1}{\varepsilon}\right)^{1/3}\, cm,\,\,
\lambda_{rest} = 0.8\, \mu m. \label{eq21}
\end{equation}

\noindent For $\varepsilon = 0.1$, which corresponds to $a/M_{BH}\simeq0.6$ in the model
of Novikov \& Thorne (1973), the ratios of $R_{\lambda}/R_H$ are equal to

\[
R_{\lambda}(0.44)/R_H = 13;\,\, R_{\lambda}(0.56)/R_H = 18;
\]
\begin{equation}
R_{\lambda}(0.80)/R_H = 29. \label{eq22}
\end{equation}

\noindent The formulae (21) for arbitrary wavelength $\lambda$ acquire the form:

\begin{equation}
R_{\lambda}=1.94\cdot
10^{14}\frac{\lambda^{4/3}}{\varepsilon^{1/3}}. \label{eq23}
\end{equation}

\noindent These estimates show that $R_{\lambda} < R_{ab}$, if the viscosity
parameter $\alpha \ge 0.1$ (see Eq.~(23)). Here $R_{ab}$
is the boundary radius between two zones of the standard accretion
disk (Shakura \& Sunyaev 1973): (a) a radiation
pressure-dominated zone and (b) a gas pressure-dominated zone. In
both zones the opacity comes from Thomson scattering. The boundary
between these zones is given by Shakura \& Sunyaev (1973) and
Pariev \& Colgate (2007):

\begin{equation}
R_{ab} = 236 R_H \left(\frac{\alpha}{0.01}\right)^{2/21}
\left(\frac{M_{BH}}{10^8 M_{\odot}}\right)^{2/21}
\left(\frac{\eta}{\varepsilon}\right)^{16/21}. \label{eq24}
\end{equation}

This means that the characteristic spatial radius of an accretion
disk $R_{\lambda}$, corresponding to observed wavelength
$\lambda$, lies for NGC 4258 in the zone (a). For the
frequently used value $\alpha =0.01$ (see Pariev \& Colgate 2007), we
have $R_{ab}=23.2 R_H$ at $\epsilon =0.1$; i.e., the inequality
$R_{\lambda}<R_{ab}$ takes place only for $\lambda =0.44\mu$m and 0.56$\mu$m.

We use the polarimetric data by Modjaz et al. (2005).
These data allow  a $1-\sigma$ upper limit of $B_{mas}\sim
30 mG$ on the radial component of the disk magnetic field at the
radius of 0.14 pc. Using the power-law radial dependence of
magnetic field (11), we obtain the following expression:

\begin{equation}
B_{mas} = B(R_{\lambda})
\left(\frac{R_{\lambda}}{R_{mas}}\right)^{5/4} = 0.018
\left(\frac{0.1}{\varepsilon}\right)^{5/12} \label{eq25}
\end{equation}

\noindent where $R_{mas} = 0.14$ pc, $B(R_{\lambda}) = 425$ G for
$\lambda =0.44\mu m$. This expression can be used for a crude estimate
of parameter $\varepsilon$.
In our case we obtain $\varepsilon \approx 0.03$. In this case,
$R_{ab}=58.2 R_H(\alpha/0.01)^{2/21}$. This means that inequality $R_{\lambda}<R_{ab}$
occurs even for value $\alpha =0.01$, mentioned above. The values $R_{\lambda}$
for $\epsilon =0.03$ are higher than values (22) to 1.5 times, i. e. they also lesser
than $R_{ab}$, if we take $\alpha =0.01$.

Finally, we can estimate the magnetic field strength $B_H$ at the
horizon radius of the black hole in NGC 4258 using the data
for $R_{\lambda}/R_{H}$ presented in expressions (22),
and for our values $B(R_{\lambda})$
for the turbulent accretion disk model (see Sect.~3.2).
Taking in Eq.~(11) $r=R_{\lambda}$, we obtain the expression:

\begin{equation}
B_H = B(R_{\lambda}) \left(\frac{R_{\lambda}}{R_H}\right)^{5/4}
(G). \label{eq26}
\end{equation}
\noindent As a result, we obtain for $\lambda=0.44\mu m, \,0.56\mu m$,
and $\lambda =0.8\mu m$ the following values  $B_H= 1.05\cdot 10^4 G,\,
6.06\cdot 10^3 G$ and $7.57\cdot 10^3 G$, respectively (at $\varepsilon=0.1$,
which corresponds to spin (Kerr) parameter $a/M_{BH}\simeq 0.6$ in the model of
Novikov \& Thorne 1973). At $\epsilon=0.03$, these values are to 1.66 times
higher. This case approximately corresponds
to $a/M_{BH}\approx -0.95$ (see Krolik 2007).

Our estimates are
slightly different as a result of different error intervals of
polarimetric data (see Eq.~(16)). Besides, some
uncertainty exists in the choice of the level of fluctuations (parameter C).
It seems for $\lambda=0.56\mu m$ that this uncertainty is less than for other
wavelengths. For this reason the estimate $B_H=6.06\cdot 10^3 G$ seems to be
preferable. As an example, we present the estimates for this effective
wavelength and other various
parameters $\varepsilon$  and $a/M_{BH}$  in Table~1.

In the estimations, presented above we take $n=5/4$ in the basic Eq.~(11).
How do we change the estimations for other values of $n$? The calculations give
the following values of $B_H$ at the event horizon of the supermassive black
hole in NGC 4258: $B_H=3\cdot 10^3$~G at $n=1$, and $B_H=5.3\cdot 10^4$~G at
$n=2$. These results indicate that the magnetic field strength of SMBH in
NGC 4258 at the event horizon should be at the level $\approx 10^3-10^4$~G.

It should be noted that our estimations do not use the values of
viscosity parameter $\alpha$. It was only shown that, for $\alpha >0.1$, the radius
$R_{\lambda}$ lies inside zone (a). Of course, our estimations depend
on parameters $\varepsilon$ and the power-law index n, which, in principle,
are to be found in a detailed model of an accretion disk.

The data in Pariev et al. (2003) are given for $r>2.5 r_g$
($r_g=2GM_{BH}/c^2$ is the gravitational radius of black hole),
i. e. slightly beyond the horizon radius $R_H$ (beyond $2.2R_H$ for $\varepsilon =0.1$,
and beyond $5R_H$ for $\varepsilon =0.03$). It seems that they used the simplified
theory, which do not ``work" near the horizon. The calculations in Pariev et al. (2003)
correspond to magnetic dominated regime; i.e., magnetic energy is greater than
radiative thermal energy. Because $R_{\lambda}$ lies inside the zone (a) (radiation
dominated zone), we derive that plasma parameter $\beta=P_{gas}/P_{magn}<1$ in model of
Pariev et al. (2003). Here $P_{gas}$ and $P_{magn}$ are gas and magnetic pressures,
respectively.

\begin{table}
\caption{\smallskip The value of $B_H$~[G] for various data of Kerr parameter
and radiative efficiency.}
\setlength{\tabcolsep}{0.1cm}
\centering
\begin{tabular}{l l r r r}
\hline
\noalign{\smallskip}
Source & $\delta_{\|};\delta_{\bot}$ & $\frac{a}{M_{BH}}=0.5$ &
$\frac{a}{M_{BH}}=0.998$ & $\frac{a}{M_{BH}}=-0.9$\\
 & & $\varepsilon =0.081$ & $\varepsilon =0.32$ & $\varepsilon =0.039$\\
\noalign{\smallskip}
\hline
\noalign{\smallskip}
NGC 4258 & 41; 3.5 & $6.6\cdot 10^3$ & $4\cdot 10^3$ & $9.6\cdot 10^3$ \\
\noalign{\smallskip}
\hline
\end{tabular}
\end{table}

\begin{table}
\caption[]{The value of linear polarization from accretion disks.}
\setlength{\tabcolsep}{0.1cm}
\centering
\begin{tabular}{l c c c c}
\hline
\noalign{\smallskip}
Source  & $\log{\frac{M_{BH}}{M_{\odot}}}$ & $\log{\frac{L_{bol}}{L_{Edd}}}$ &
$p$ [\%] & $p$ [\%] \\
 &  &  & $\frac{a}{M_{BH}}=0.5$ & $\frac{a}{M_{BH}}=0.998$ \\
 &  &  & $\varepsilon =0.081$   & $\varepsilon =0.32$ \\
\noalign{\smallskip}
\hline
\noalign{\smallskip}
BLRG      & 8.64 & $-$1.5  & 0.06 & 0.057 \\
NLRG      & 8.14 & $-$2.92 & 0.09 & 0.067 \\
SSRQ      & 9.29 & $-$0.90 & 0.05 & 0.057 \\
FSRQ      & 9.01 & 0.44    & 0.94 & 0.042 \\
Seyfert 1 & 7.23 & $-$0.59 & 1.40 & 0.036 \\
\noalign{\smallskip}
\hline
\end{tabular}
\end{table}

\section{Magnetic coupling process in AGN and QSO: testing by
continuum polarization}

Li (2002), Wang et al. (2002, 2003),
Zhang et al. (2005) studied the
magnetic coupling process (MC) as an effective mechanism for
transferring energy and angular momentum from a rotating black hole
to its surrounding accretion disk. This process can be considered
as one of the variants of the Blanford-Znajek (BZ) process
(Blanford \& Znajek 1977; Blanford \& Payne
1982). It is assumed that the disk is stable,
perfectly conducting, thin, and Keplerian. The magnetic field is
assumed to be constant on the black hole horizon and to vary as a
power law with the radius of the accretion disk.

Since the magnetic field on the horizon $B_H$ is brought and held
by its surrounding magnetized matter of a disk,
the some relation must exist between the magnetic field strength and
accretion disk and, finally, the bolometric luminosity
of AGN (see Ma et al. 2007). This relation takes a form

\begin{equation}
B_H(G) = \frac{k^{1/2} (2 L_{bol}/\varepsilon c)^{1/2}}{R_H}.
\label{eq27}
\end{equation}

\noindent Here $L_{bol} = \varepsilon \dot{M} c^2$, $R_H$ is the horizon
radius (see Eq.(12)), $\dot{M}$ is the
accretion rate, $\varepsilon$ the radiative efficiency
(calculated, for example, by Novikov \& Thorne 1973;
Krolik 2007; Shapiro 2007, see also Sect.~2.2). Coefficient
$k$ is the inverse plasma parameter $\beta =
P_{gas}/P_{magn}$, where $P_{gas}$ and $P_{magn}$ are gas and
magnetic pressures, respectively. We shall consider
the case $\beta = 1$ ($k = 1$) in future, i.e. the equipartition case.

Then Eq.~(27) is transformed into

\begin{equation}
B_H(G) = \left(\frac{M_{\odot}}{M_{BH}}\right)^{1/2}
\left(\frac{\eta}{\varepsilon}\right)^{1/2} \frac{6.2\cdot
10^8}{1+\sqrt{1-(a/M_{BH})^2}}, \label{eq28}
\end{equation}

\noindent where $a/M_{BH}$ is the Kerr parameter, $\eta = L_{bol}
/ L_{Edd}$ and $L_{Edd} = 1.3\cdot 10^{38} (M_{BH}/M_{\odot})$(erg s$^{-1}$) is
the Eddington luminosity.

We next calculate the expected polarization value of radiation in a
number of specific AGNs, taking the effect of Faraday
depolarization into account, as considered above. In mechanisms of magnetic coupling,
we mainly consider the case where magnetic field is supposed directed
along the normal to the accretion disk (see, for example,
Wang et al. 2002, Ma et al. 2007). Our estimates of the magnetic field in NGC 4258 (see
Sect.~3.2) show that the vertical magnetic field is much larger than the perpendicular one.
For these reasons, we also consider only ${\bf B}_{\|}$ fields as
first approximation, so we have to determine
$B_{\|}=B(R_{\lambda})$ from Eq.~(25). The values $R_H$ and $R_{\lambda}$ are presented
in Eqs.(12) and (14), and $B_H$ - in Eq.~(28). As a result, we obtain the
following formula for dimensionless parameter $\delta_{\|}$:

\[
\delta_{\|}=0.8\lambda^2(\mu m)B_{\|}(G)\simeq 474\lambda^{1/3}(\mu m)
\left(\frac{\eta}{\varepsilon}\right)^{1/12}\times
\]
\begin{equation}
\left(\frac{M_{BH}}{M_{\odot}}\right)^{-1/12}
\left(1+\sqrt{1-(\frac{a}{M_{BH}})^2}\right)^{1/4}. \label{eq29}
\end{equation}
\noindent The estimations of polarization degree $p(\lambda)$ can be obtained
from Eq.~(9).

A systematic analysis of a large sample of AGN
available in the BeppoSAX public archive was performed by
Grandi et al. (2006). Their sample includes AGN of
various types. Narrow line radio galaxies (NLRG), broad line radio
galaxies (BLRG), steep spectrum radio quasars (SSRQ) and flat
spectrum radio quasars (FSRQ) (see Table~6 from their paper, where
the values $M_{BH}/M_{\odot}$ and $\eta=L_{bol}/L_{edd}$ are presented.

The results of our calculations of the dimensionless
depolarization parameter $\delta_{\|}$ are

\[
BLRG:\, \delta_{\|} \simeq 79; 60;\,\,\,\, NGRG:\, \delta_{\|} \simeq 67; 57,
\]

\[
SSRQ:\, \delta_{\|} \simeq 79; 60;\,\,\,\, FSRQ:\, \delta_{\|} \simeq 108; 82,
\]

\begin{equation}
Seyfert\, 1:\, \delta_{\|} \simeq 125; 95, \label{eq30},
\end{equation}

\noindent where the first numbers correspond to $\varepsilon=0.081$ and $a/M_{BH}=0.5$,
and the second ones correspond to $\varepsilon=0.32$, $a/M_{BH}=0.998$.

The estimations of possible $p(\mu)$ for the inclination angle of the
accretion disk $i=60^{\circ}$ ($\mu =\cos i, p(0,\mu)=2.25\%$) are presented in Table~2.
Because all values of $\delta_{\|}\gg 1$, we can use the approximate formula
$p(\mu )\simeq p(0,\mu)/\delta_{\|}\mu $. This formula is valid up to $i=87^{\circ}$,
so for $i=85^{\circ}$ ($p(0,\mu)\simeq 7.8\%, \mu = 0.087$) the polarization degree
is to 20 times higher than values presented in Table ~2
($p(\mu=0.087)=(p(0,0.087)/p(0,0.5))(0.5/0.087)\simeq 20$). For other inclination angles
$i$, the calculations are analogous. We stress that polarization observations of sources,
presented in Table~2, do not exist up to now, so we present only possible values of
polarization degrees. This procedure, presented below, can be used to estimate
the source parameters if the polarization data are available.

The limiting case $a/M_{BH}=0$ corresponds to $\varepsilon =0.057$; i.e.,
the predicted polarizations differ only slightly from the
presented values for $a/M_{BH}=0.5$, so for the source NGC 4258 instead
of $6.6\cdot 10^3$~G, corresponding to $a/M_{BH}=0.5$ (see Table~1),
we obtain $7.6\cdot 10^3$~G for limiting case $a/M_{BH}=0$.

\section{The wavelength dependence of polarization
of AGN and accretion disk models}

Polarization in AGNs can be intrinsic or extrinsic. Light scattering by
a nonspherical distribution of electrons near the central engine
of AGN is a basic intrinsic polarization mechanism;
namely, an accretion disk is a typical example of a nonspherical
distribution. Scattering by magnetically aligned dust grains in
the interstellar medium of galaxies is the typical example of an
extrinsic situation.
The very important feature characterizing the polarized radiation
from a magnetized accretion disk is the wavelength dependence of
polarization degree that is very different from that of
Thomson's scattering.

For a strong magnetic field strength, when $\delta_{\| ,
\bot}\gg 1$ the simple asymptotic formulas follow from
Eqs.(9) and (10):

\[
p(\lambda) \approx
\frac{p(0,\mu)}{\delta_{\|}\mu}\sim \frac{p(0,\mu)}{B_{\|}\lambda^2},
\]
\begin{equation}
p(\lambda)\approx \frac{p(0,\mu)}{\delta_{\bot}\sqrt{1-\mu^2}}
\sim\frac{p(0,\mu)}{B_{\bot}\lambda^2}. \label{eq31}
\end{equation}

\noindent For turbulent atmospheres, with  the existence of chaotic magnetic
field component $B'$, the corresponding asymptotic formulae follow from
Eqs.~(20). If the parameter of turbulence $C\gg \delta \gg 1$,
then $p(\lambda)\sim 1/\lambda^4$, if the level of magnetic fluctuations
$\langle B'^2\rangle$ is constant (see Sect.~3.2).
Below we mainly consider the limiting spectra
($\delta_{\|,\bot}\gg 1)$ only for nonturbulent atmospheres. For turbulent
atmospheres with a high value of $C$, the corresponding spectra are to be
divided to $\lambda^2$.

The wavelength dependencies of the radiation flux and its
polarization essentially stem from the radial distribution of the
temperature in an accretion disk. For a standard accretion disk
(Shakura \& Sunyaev, 1973), radial dependence of the temperature takes
the form: $T_e\sim r^{-3/4}$. To get the integrated spectrum from the disk,
we add up all of the Planck curves from each radius. If $T_e\sim r^{-s}$
then the radiation flux (see Pringle \& Rees 1972,
Shakura \& Syunyaev 1973, Gaskell 2008):

\begin{equation}
F_{\nu}\sim \nu^{3-2/s}. \label{eq32}
\end{equation}

\noindent For a standard accretion disk $(s=3/4$), the flux (30) takes
the known form $F_{\nu}\sim \nu^{1/3}$.

Substitution of $T_e\sim r^{-s}$
into formula (13) leads to relation $R_{\lambda} \sim \lambda_e^{1/s}$
between the characteristic scale $R_{\lambda}$ and the effective wavelength of
polarimetric observations. According to
expression (11) we obtain $B_{\|,\bot}(R_{\lambda})\sim \lambda^{-n/s}$.
As a result, from Eqs.~(31) and (11), we obtain the next wavelength
dependence for the case of a strong Faraday depolarization:

\begin{equation}
p(\lambda)\approx \frac{p(0,\mu)}{B_{\|,\bot}\,\lambda^2}\sim
\lambda^{n/s - 2}. \label{eq33}
\end{equation}
\noindent As a result, the degree of observed polarization $p(\lambda)$ depends on both
power-law indices $n$ and $s$. This comes from the depolarization effect
of Faraday rotation of the polarization plane. In the absence of this effect, the polarization
degree, independent of radiation flux $F_{\lambda}$, depends only on the inclination
angle of the accretion disk. Of course, the usual Stokes parameters $I,Q$, and $U$ give
the information on the radial distribution of the effective temperature (e.g. Eq.~(32)).

For distribution of magnetic field in a standard
accretion disk ($s=3/4$) with $n=3/2$ ($B_{\|, \bot}\sim R^{-3/2}$),
the polarization does not depend on the wavelength. For a standard accretion disk
with $n=5/4$, mostly elaborated by Pariev et al. (2003), the wavelength dependence
of the polarization is quite weak $p(\lambda)\sim \lambda^{-1/3}$. In this case
the wavelength dependence of the Stokes polarized flux takes the
form $p(\nu)F_{\nu}\sim \nu^{2/3}$.
In the general case the formulae (32) and (33) give rise to relation

\begin{equation}
p(\nu)\,F_{\nu}\sim \nu^{5-\frac{n+2}{s}}. \label{eq34}
\end{equation}

The presented formulas allow us to test the various models of an accretion disk
using the data of the wavelength dependence of polarization of AGN
and quasars.

For the turbulent accretion disk with a high value of parameter $C$,
formula (34) transforms to the expression

\begin{equation}
p(\nu)\,F_{\nu}\sim \nu^{7-\frac{n+2}{s}}. \label{eq35}
\end{equation}
\noindent The position angle values $\chi$, used in our
formulae, denote the angles between the accretion disk plane and the observed
direction of electric field wave oscillations. If it is impossible to estimate
the disk inclination, we cannot obtain the theoretical $\chi$ from the
observed position angle. The observed spectra of polarization degree $p_{obs}(\lambda)$
and the differences of position angle $\chi_{obs}(\lambda_i)-\chi_{obs}(\lambda_j)$
are to coincide with the corresponding theoretical
values $p(\lambda)$ and the differences in position angles
$\chi(\lambda_i)-\chi(\lambda_j)$ for different wavelengths $\lambda_i$
and $\lambda_j$. These conditions make the estimation of magnetic field and unknown
disk's inclination angle more precise than for the single wavelength observations.

\subsection{Discussion of some observational data}

Webb et al. (1993) and Impey et al. (1995) present the data
of measurement UBVRI polarizations of a sample of AGNs and QSOs.
The position angle appears to be wavelength-independent,
suggesting that the polarization in a given object originates in a
single physical process. In many cases the percentage of
polarization increases with frequency. Authors have compared the
polarized fluxes with the predictions of competing models of
polarization in AGNs: synchrotron emission, scattering from
electrons or different types of dust grains, and electron
scattering in an accretion flow. In nine sources from this sample,
the polarization seems to be the result of dust grain
scattering. A number of these sources (NGC 4151, Mrk 509, NGC
5548, Mrk 290) has best characteristics of model due to electron
scattering in an accretion disk or torus. It is interesting that,
for these four sources, the slopes of $F_{\nu}\sim \nu^{\gamma}$ are
close ($\gamma \simeq -1.37, -0.9, -0.86, -0.9$, respectively;
see Webb et al. (1993). According to formula (32), we obtain the
corresponding values of parameter $s=0.46, 0.47, 0.52, 0.51$; i.e.,
they are near $s=0.5$.

The general shape of the polarization spectrum with parameter
''q'' was determined by a least-squares fit proportional to
$p(\nu)F_{\nu}=A\nu^{q}$. It appears that the slope of the
wavelength dependence of polarized flux $q$ varies widely, between
$-2$ and $+1$, with a typical uncertainty of 0.3.

What values of parameter $n$ corresponds to limit values of
$q$, if we accept $s=1/2$ using formulas (34) and (35)?
For nonturbulent accretion disk (Eq.~(34)) we obtain

\begin{equation}
n=\frac{1-q}{2}. \label{eq36}
\end{equation}
\noindent This formula gives $n=1.5$ for $q=-2$, $n=1$ for $q=-1$, $n=0.5$ for $q=0$,
and $n=0$ for $q=1$.
Remember (see Pariev et al. 2003) that accretion disk with
decreasing gas density at large distances $r$ corresponds
to $n>1$. It seems the self-consistent models of Pariev et al. (2003) allow
the values of $n\simeq 1-2$. The high values with $n>2$ correspond to a
very sharp decrease the magnetic field inside the accretion disk.
That is why we investigate the cases $1<n<2$ as physically acceptable
according to the results of Pariev et al. (2003). From Eq.~(32) we see
that $p_{\lambda}\sim \lambda^{2(n-1)}$ at $s=1/2$. This means that the degree
of polarization that diminishes with the increasing  $\lambda $ cannot be
explained inside the physical limitations, assumed above ($n>1$ corresponds
to increasing polarization degree with $\lambda$).

For this reason we investigate what gives the very high magnetic turbulence.
For this case, formula (32) transforms to expression $p(\lambda)\sim \lambda^{n/s-4}$.
It leads to $p(\lambda)\sim \lambda^{2(n-2)}$ at $s=1/2$. As a result, our physically
acceptable restriction $1<n<2$ allows us, in principle, to explain the decreasing
degree of polarization. Taking the definition $p(\nu)F_{\nu}\sim \nu^q$ into account,
we derive from Eq.~(34) the following relation

\begin{equation}
n=(7-q)s-2. \label{eq37}
\end{equation}
\noindent  For $s=0.5$ this expression gives $n=1$ for $q=1$, and $n=1.5$ for $q=0$.
For $q=-2$ this formula gives $n=5/2$, which characterizes a very sharp decrease in
magnetic field inside the accretion disk.

As an example, we now consider the continuum radiation of source NGC4151. According
to Webb et al. (1993),
we have $\gamma\simeq -1.37$, and from Schmidt \& Miller (1980)
the parameter $q$ is equal to $q\simeq -0.33$ (see Fig.~3 for the
value $p(\nu)F_{\nu}$ in the interval $\lambda =(0.3 - 0.94)\mu$m).
From Eq.~(31) we obtain $s=2/(3-\gamma)$, which gives in
this case $s\simeq 0.46$. Substitution of the values $q=-0.33$ and $s=0.46$ to Eq.~(37)
gives $n\simeq 1.37$. In nonturbulent case we have $n\simeq 0.45$, which
corresponds to the increase in gas density with growing distance from the
nucleus (see Pariev et al. 2003).  In
our case, when we know the observed spectra $F_{\nu}\sim \nu^{\gamma}$, and
$p(\nu)F_{\nu}\sim \nu^q$, both Eqs.~(34) and (35) give rise to the same
expression $p(\lambda)\sim \lambda^{\gamma - q}\simeq \lambda^{-1.04}$. Thompson
et al. (1979, see Fig.~3) present the polarization degree $p(\lambda)\%$ in the same
interval $\lambda =(0.3 - 1)\mu$m. The data have a rather wide distribution.
Using our dependence, we can approximate the polarization degree
by the formula $p(\lambda)\simeq -0.5+0.7\cdot\lambda^{-1.04}(\mu$m),
which fairly satisfactorily describes the mean value of the presented data.

\section{Conclusions}

We have presented the method for estimating the magnetic field
strength at the event horizon of a supermassive black hole through
the polarization of accreting disk emission. The polarized
radiation arises from the scattering of emission light by
electrons in a magnetized accretion disk. Due to Faraday rotation of
the polarization plane, the resulting polarization degree differs essentially
from the classical Thomson case, because the wavelength dependence of the polarization
degree appears. This feature means that the magnetic field strength at the
event horizon of a black hole can be estimated from polarimetric
observations in the optical range.

For estimating observed polarization,
we use the azimuthally averaged asymptotic expressions of the Stokes parameters
for outgoing radiation, assuming that the accretion disk is optically thick and that
the Milne problem takes place. Using these formulae, we discuss the wavelength
dependence of the observed spectra of polarization degree. We also consider
turbulent accretion disks when the magnetic field possess both regular
component and chaotic one. Since the polarization spectrum of
scattered radiation strongly depends on the accretion disk model, our
results can be used to construct a  realistic physical model of
the AGN environment.

The estimates of the magnetic field strength of
supermassive black holes in NGC 4258 are presented. We also found for the source
NGC 4151 that the observed polarization degree spectrum can be satisfactorily
explained by our mechanism with the high level of turbulent magnetic
field. In this case the power-law regular magnetic field is $B\sim r^{-1.37}$
and $T_e \sim r^{-0.46}$.
In the cases where the observed polarization is the result of various mechanisms,
such as light scattering in an accretion disk and a jet, our method can be considered
as important part of the problem as a whole.

\section*{Acknowledgements}
This research was supported by the RFBR (project No.
07-02-00535a), the program of Prezidium of RAS ''Origin and
Evolution of Stars and Galaxies'', the program of the Department
of Physical Sciences of RAS ''Extended Objects in the Universe''
and by the Grant from President of the Russian Federation ''The
Basic Scientific Schools'' NS-6110.2008.2. M.Yu. Piotrovich
acknowledges the Council of Grants of the President of the Russian
Federation for Young Scientists, grant No. 4101.2008.2.

We are very grateful to the anonymous referee for many remarks that allowed us
to clarify the paper considerably.

\end{document}